\documentclass{article}
\usepackage{amsmath, amsthm}

\usepackage{PRIMEarxiv}

\usepackage[utf8]{inputenc} 
\usepackage[T1]{fontenc}    
\usepackage{hyperref}       
\usepackage{url}            
\usepackage{booktabs}       
\usepackage{amsfonts}       
\usepackage{nicefrac}       
\usepackage{microtype}      
\usepackage{lipsum}
\usepackage{fancyhdr}       
\usepackage{graphicx}       
\graphicspath{{media/}}     

\usepackage{amsmath}
\usepackage{amssymb}
\usepackage{hyperref}
\usepackage{algorithm}
\usepackage{listings}
\usepackage{thmtools}
\usepackage{thm-restate}
\usepackage{algpseudocode}
\usepackage{bbm}
\usepackage{verbatim}
\usepackage{xcolor}
\newcommand{\poly}{\textnormal{poly}}
\newcommand{\dominik}[1]{{\color{blue} {Dominik says: #1}}}

\newcommand{\toappendix}[1]{{\color{purple} {Material marked for appendix hidden here.}}}

\newcommand{\vbl}{\textnormal{vbl}}
\DeclareMathOperator*{\E}{\mathop{\mathbb{E}}}
\renewcommand{\root}{\textnormal{root}}
\newcommand{\clauselabel}{\textnormal{clauselabel}}
\newcommand{\varlabel}{\textnormal{varlabel}}
\newcommand{\plaus}{\textnormal{Plaus}}

\newcommand{\evplausimp}{\textnormal{EvPlausImp}}
\newcommand{\cut}{\textnormal{Cut}}
\newcommand{\impcut}{\textnormal{ImpCut}}
\newcommand{\localimpcut}{\textnormal{LocalImpCut}}
\newcommand{\error}{\textnormal{error}}
\newcommand{\abamo}{\textnormal{abamo}}

\newcommand{\costspelledout}{\E\left[\log_2( A_x) \right]}

\newcommand{\costimpspelledout}{\E\left[\log_2( \max(1 + \eligible_x, A^{\imp}_x)) \right]}

        \newtheorem{theorem}{Theorem}
		\newtheorem{lemma}[theorem]{Lemma}
		\newtheorem{corollary}[theorem]{Corollary}
		\newtheorem{proposition}[theorem]{Proposition}
		
		\newtheorem{definition}[theorem]{Definition}
		
		\newtheorem{observation}[theorem]{Observation}
		
\title{Impatient PPSZ --- a Faster algorithm for CSP}

\author{
  Shibo Li  \\
  Shanghai Jiao Tong University \\
  Shanghai\\
  \texttt{ShiboLi@sjtu.edu.cn}
   \And
  Dominik Scheder \\
  Shanghai Jiao Tong University \\
  Shanghai\\
  \texttt{dominik@cs.sjtu.edu.cn} \\
}

\begin{document}
\maketitle

\begin{abstract}
PPSZ is the fastest known algorithm for $(d,k)$-CSP
problems, for most values of $d$ and $k$. 
It goes through the variables in random order
and sets each variable randomly to one of the $d$ colors,
excluding those colors that can be ruled out by 
looking at few constraints at a time.

We propose and analyze a modification of PPSZ:
whenever all but 2 colors can be ruled out 
for some variable, immediately set that variable
randomly to one of the remaining colors. We show
that our new ``impatient PPSZ'' outperforms PPSZ exponentially
for all $k$ and all $d \geq 3$ on formulas 
with a unique satisfying assignment.
\end{abstract}


\section{Introduction}
\label{sec:Introduction}
A Constraint Satisfaction Problem, or CSP for short, consists of a finite a set of variables $x_1,\dots, x_n$, a domain $[d] := \{1,\dots,d\}$ of potential values, and a set of constraints. A constraint
is of the form $(x_{i_1}, \dots, x_{i_k}) \in S$, where $S \subseteq [d]^k$. In analogy to CNF-SAT, we assume
in this paper that $|S| = d^k-1$, i.e., 
all but one 
possible assignments satisfy the constraint.
We speak of a {\em $(d,k)$-CSP} if 
all constraints are over $k$ variables.
In a slight abuse of notation, we also use $(d,k)$-CSP to denote
the associated {\em decision problem}: 
is there a way to assign values in $[d]$
to the variables that satisfies all constraints?
This is NP-complete except when
$d=1$ or $k=1$ or $k=d=2$, so researchers focus
on finding {\em moderately exponential algorithms}:
algorithms of running time $c^n$ for $c < d$. Examples include
Beigel and Eppstein's 
randomized algorithm for $(d,2)$-CSP with running time $O((0.4518d)^n)$~\cite{beigel20053};Sch\"oning' s random walk algorithm of running time $O^*((\frac{d(k-1)}{k})^n)$~\cite{schoning1999probabilistic}; Paturi, Pudl\'ak, and Zane encoding-based randomizedalgorithm called PPZ~\cite{paturi1997satisfiability} for
 $k$-SAT (i.e., $d=2$), which runs in time 
$O(2^{(1-1/k)n})$.  Paturi, Pudl\'ak, Saks and Zane~\cite{paturi2005improved} improved PPZ 
by introducing  a pre-processing step using small-width resolution. 
Both PPZ and PPSZ can be easily modified to work for $(d,k)$-CSP as well,
as done by Scheder~\cite{scheder2010ppz} for PPZ and
Hertli et al. \cite{hertli2016ppsz} for PPSZ.
In both cases, several subtleties and technical difficulties arise,
which are not present for $k$-SAT.
Furthermore, \cite{hertli2016ppsz} is the currently fastest algorithm for $(d,k)$-CSP when $k \geq 4$. 

\subsection{The PPSZ Algorithm}

Let us give an informal description of PPSZ, first for SAT, 
and then for CSP. In either case, it chooses a random ordering
$\pi$ on the variables $x_1,\dots,x_n$. Then it goes through the
variables one by one, in the order of $\pi$;
when processing $x_i$, it fixed $x_i$ randomly to 
\texttt{true} or \texttt{false}, unless the correct value can be inferred by 
a set of up to $D$ clauses (in which case we say 
$x_i$ has been inferred by $D$-implication). 
For CSP, the only difference
is that when processing $x_i$, it checks (with brute force)
for which colors $c \in [d]$ the statement $[x_i \ne c]$ can be 
inferred by a set of up to $D$ constraints; 
if so, we say $[x_i \ne c]$ is {\em $D$-implied}, and color $c$
is obviously ruled out. It then fixes $x_i$ randomly to one 
of the colors not yet ruled out (or declares failure if all
colors have been ruled out).\\

\textbf{Unique-SAT versus general-SAT}. A peculiar feature of PPSZ,
as analyzed in the seminal paper~\cite{paturi2005improved},  
is that it performs better if the input instance $F$ has a {\em unique} satisfying assignment. Certain properties, such as the existence of critical clause trees, break down once $F$ has multiple solutions. In~\cite{paturi2005improved}, the authors proposed a clever but technical workaround, which incurred an exponential overhead for $k=3,4$. In his 2011 breakthrough paper, Hertli~\cite{hertli2011} showed that this peculiarity is in fact an artifact of the analysis, and gave a very abstract and high-level proof that PPSZ on formulas with many solutions is indeed no worse. His proof was later simplified by Scheder and Steinberger~\cite{SchederSteinberger}. The proofs in
\cite{hertli2011} and \cite{SchederSteinberger} work only
provided
that the internal machinery of the PPSZ algorithm (e.g., checking $D$-implication) is ``not too good''. Curiously, in~\cite{hertli2016ppsz} it turned out
that, for $k=2,3$ and certain values of $d$, 
the PPSZ machinery is indeed ``too good'', and consequently
their time complexity for the general case (multiple solutions)
is worse than for the unique case (exactly one solution).
For formulas with a unique solution, their analysis gives
the best known running time
for all $d,k$ except for $k=2$ and $d \in \{3,4\}$.
\\

\textbf{Improvements to PPSZ for $k$-SAT.} Two 
recent results improve PPSZ. Hansen, Kaplan, Zamir,
and Zwick~\cite{HKZZ} define a {\em biased}
 version of PPSZ and show that it achieves an improvement
 for all $k \geq 3$. Scheder~\cite{scheder2021eccc}
 shows that PPSZ 
 itself performs exponentially better than 
 in the analysis of~\cite{paturi2005improved}.
We would not be surprised if both improvements carry
over to $(d,k)$-CSP, although to our knowledge,
this has not been analyzed so far.
The improvement presented in this work is of different quality: it is not a generalization of some 
idea for $k$-SAT; in fact, the main idea only makes
sense for $d \geq 3$ and thus is particular to $(d,k)$-CSP problems.\\

\textbf{The time complexity of PPSZ for Unique $(d,k)$-CSP.}
A main result of~\cite{hertli2016ppsz} is that 
PPSZ solves Unique $(d,k)$-CSP in time $O\left(2^{S_{d,k} \, n + o(n)} \right)$, where $S_{d,k}$ is defined by the 
following random experiment: let $T^{\infty}$ be 
the infinite rooted
tree in which each node on even depth (which includes
the root at depth $0$) has $k-1$ children and 
every node on odd depth has $d-1$ children.
Let $T_1, \dots, T_{d-1}$ be disjoint copies of 
$T^{\infty}$, 
sample $p \in [0,1]$ uniformly, and delete every
odd-level node with probability $p$, independently.
Let $J_c$ be the indicator variable that is $1$ if the
root of $T_c$ is contained in an infinite component
after this deletion step. Then
\begin{align}
\label{eq-definition-Sdk}
    S_{d,k} := \E[\log_2 (J_1 + \cdots + J_{d-1} + 1)] \ .
\end{align}

\subsection{Our Contribution}\label{OurContribution}

In this work, we focus
on the case that $F$ has a unique satisfying assignment
$\alpha^*$, without loss of generality  $\alpha^* := (d,d,\dots,d)$.
The idea behind our improvement is as follows: 
suppose $x,y,z$ are 
variables appearing in the order $y,x,z$ in 
$\pi$. Focus on the
point in time when PPSZ processes $x$, and assume every
assignment prior to $x$ has been correct. For example, 
the variable $y$ has already been replaced by the constant $d$.
In other words, when PPSZ tries to infer statements like 
$[x \ne c]$ from small sets of constraints, it can use 
the information $[y=d]$. It cannot use $[z=d]$, however. Or can it? Maybe PPSZ can already infer $[z\ne 1], \dots, [z \ne d-1]$; in this case, it can also infer $[z = d]$, and it would be safe to fix $z$ to $d$. Let us propose the following rule:

\begin{quotation}
   \textbf{Rule of One.}
   Whenever $[z = c]$
   can be inferred by $D$-implication, fix $z$ to $c$.
\end{quotation}

This rule is ``uncontroversial'' in the sense that it will
never make a mistake. However, the reader who is familiar with
the literature about PPSZ, in particular with its original
version using small-width resolution, will notice
that resolution implicitly implements the above rule. 
We propose the following more aggressive rule:

\begin{quotation}
   \textbf{Rule of Two.}
   Whenever $[z = c_1 \vee z = c_2]$ 
   can be inferred by $D$-implication, i.e., if all 
   but 2 colors can be ruled out, pick $c \in \{c_1, c_2\}$ 
   arbitrarily and fix $z$ to $c$.
\end{quotation}

Obviously, this rule can introduce mistakes. On the plus side,
it might be very unlikely that the range of plausible (i.e., not ruled out) colors for $z$ further decreases from $2$ to $1$. Better to bite the bullet now, decide on a value for $z$, hope that 
it is correct, and use that information for subsequent $D$-implications. For example, it might be that using the information  $[z=d]$
lets us rule out additional colors for $x$,
the variable currently being processed by PPSZ. 
Unfortunately, this rule does more bad than good: consider the variables coming towards the very end 
of $\pi$. For each of them, it is very like that
all but one color can be ruled out; thus,
PPSZ would set them correctly with high probability;
using our Rule of Two, this probability would go 
down from (almost) 1 to (roughly) 1/2 since 
we decide on a value once only two values are left.
We propose a less impatient rule:

\begin{quotation}
   \textbf{Conservative Rule of Two.} 
   Apply the Rule of Two only to variables $z$
   that are among the first $\theta n$ in $\pi$;
   don't apply it to the last $(1-\theta) n$ variables.
\end{quotation}

We will show that for those early variables,
it is extremely unlikely that the set of plausible
colors gets narrowed down to only one color;
and that it is somewhat more likely that
the Rule of Two helps us rule out one 
additional colors for a variable. In particular,
we prove the following theorem:

\begin{theorem}
   \label{theorem-main}
   For every $d \geq 3$ and $k \geq 2$, there is some
   $\epsilon > 0$ and a randomized algorithm solving
   $(d,k)$-CSP in time $2^{n( S_{d,k} - \epsilon)} \poly(n)$.
\end{theorem}

\subsection{Notation}

Let $V = \{x_1,\dots,x_n\}$ be a set of variables 
and $[d] = \{1,\dots,d\}$ be the set of possible colors.
A {\em literal} is an expression $(x \ne c)$, 
where $x \in V, c \in [d]$. A {\em clause} is a 
disjunction of literals: 
$(v_1 \neq c_1 \vee v_2 \neq c_2 \vee ... \vee v_k \neq c_k)$. A $(d,k)$-CSP is a conjunction of clauses 
of size $k$ each. 
An assignment $\alpha$ is a function 
$V \rightarrow [d]$. It satisfies a literal $(x \neq c)$ 
if $\alpha(v) \neq c$; it satisfies a clause if it satisfies at least one literal therein; it satisfies a $(d,k)$-CSP 
$F$ if it satisfies all clauses in $F$. 
If $V' \subseteq V$ and
 $\alpha: V' \rightarrow [d]$, we call $\alpha$ a {\em partial assignment}; $\vbl(\alpha)$ denotes its domain,
 i.e., $V'$.
 $F^{[\alpha]}$ is the simplified formula after setting all variables in $V'$ according to $\alpha$.  We will write partial assignments like this: $[x \mapsto 2, y \mapsto 3,...]$ and therefore
 $F^{[x \mapsto 2]}$ will denote the formula after replacing $x$ with $2$. 
 For a clause $C$ and a $(d,k)$-CSP $F$, $\vbl(C)$ and 
 $\vbl(F)$ denote the sets of variables in $C$ and $F$,
 respectively. For a rooted tree $T$ and a node $v$ therein, the {\em subtree of $T$ rooted at $v$}
is the tree containing $v$ (as root) and all its descendants. We use the notation
$[\mathtt{statement}]$, which evaluate to $1$ if $\mathtt{statement}$ holds, and to $0$ otherwise.

\subsection{PPSZ and impatient PPSZ }
\begin{definition}[$D$-implication~\cite{hertli2016ppsz}]
Let $F$ be a  $(d,k)$-CSP formula and $u$ be a literal of $F$. We say {\em $F$ implies $u$} and write $F \vDash u$
if all assignment satisfying $F$ also satisfy $u$. We say $F$ {\em $D$-implies $u$} and  write $F \vDash_D u$
if there is some $G \subseteq F$ with $|G| \le D$ and $G \vDash u$.
\end{definition}
 For the rest 
of the paper,  $D = D(n)$ will
 be some slowly growing function in $n$, so 
 $F \vDash_D u$ can be checked in time
$\emph{O}(|F|^D \poly(n))$, which is subexponential in $n$.

\begin{definition}[Plausible values]
   Let $F$ be a $(d,k)$-CSP formula and $x$ a variable.
   We say color $c \in [d]$ is $D$-plausible
   for $x$ in $F$
   if $F$ does not $D$-imply $(x \ne c)$.
   Let $\plaus(x,F,D)$ denote the set of all colors
that are $D$-plausible for $x$.
We will drop the parameter $D$ if it is understood
from the context.
\end{definition}


\begin{algorithm}[htb]
	\caption{PPSZ algorithm}
	\label{PPSZ}
	\begin{algorithmic}[1] 
		\Procedure{PPSZ}{$F, \pi$}  
		\State $\alpha \gets$ the empty assignment
        \For{$x \in \vbl(F)$ in the order of $\pi$}
		    \State choose $c \in \plaus(x,F^{[\alpha]})$ uniformly at random
		    \label{line-ppsz-pick}
		    \State $\alpha := \alpha \cup [x \mapsto c]$
		 \EndFor
		\State \textbf{return} $\alpha$ if it satisfies $F$, else failure
		\EndProcedure
	\end{algorithmic}
\end{algorithm}

Note that our code   specifies $\pi$
as an explicit input parameter; it is the responsibility
of the ``user'' to make sure ${\rm PPSZ}(F, \pi)$ is 
called with a random $\pi$; furthermore,
we implicitly assume that PPSZ declares failure 
if the set $\plaus(x, F^{[\alpha]})$ in Line~\ref{line-ppsz-pick} is empty. 
\newcommand{\eligible}{\mathbb{I}}
From now on, we view $\pi$ not as a permutation of the variables
but as a function $V \rightarrow [0,1]$; note
that if $\pi: V \rightarrow [0,1]$ is sampled 
uniformly at random, it will be an injection
with probability $1$; sorting $V$ in ascending
order by their $\pi$-value will give 
a permutation of $V$. Additionally,
we fix two parameters $\theta$ (to be determined
later) and $c := 2 - \log_2(3)$, and mark
every variable $x$ as {\em eligible for impatient
assignment} as follows:
\begin{definition}[Eligible for impatient assignemnt]
For each variable $x$, define
$\eligible_x \in \{0,1\}$ as 
follows.
(1) If $\pi(x) \geq \theta$, set $\eligible_x := 0$;
(2) if $\pi(x) < \theta$, set $\eligible_x := 1$ with 
probability $c$ and to $0$ with probability $1-c$,
independently of all other choices. 
If $\eligible_x =1$ we say $x$ is 
{\em eligible for impatient assignment}.
\end{definition}

\begin{algorithm}[htb]
	\caption{Impatient PPSZ}
	\label{ImpatientPPSZ}
	\begin{algorithmic}[1] 
		\Procedure{ImpatientPPSZ}{$F, \pi$}  
		\State $\alpha :=$ the empty assignment
		\For{$x \in \vbl(F)$ in ascending order of $\pi$}\label{impForLoop}		
		\While{$\exists y \in \vbl(F) \setminus \vbl(\alpha)$ with  $\eligible_y = 1
		\textnormal{ and } |\plaus(y,F^{[\alpha])}|\le 2$ }
		\label{line-impatient-ppsz-while}
		    \State choose $c \in \plaus(y,F^{[\alpha]})$ uniformly at random \label{impchoosevalue}  
		    \State $\alpha  := \alpha \cup [y \mapsto c]$
		    \label{impAssigned}
		    \EndWhile
		    \label{line-after-while}
		    \If{$x \not \in \vbl(\alpha)$}
		        \State choose $c \in \plaus(x,F^{[\alpha]})$ uniformly at random \label{normalchoosevalue}
		        \State $\alpha := \alpha \cup [x \mapsto c]$
		        \label{normalAssigned}
		    \EndIf 
		\EndFor\label{endFor}
		\State \textbf{return} $\alpha$ if it satisfies $F$, else failure
		\EndProcedure
	\end{algorithmic}
\end{algorithm}

\section{Analysis of ImpatientPPSZ}

\newcommand{\imp}{\textnormal{imp}}
\textbf{Notation for sets of variables coming before variable $x$: $V_x$ and $V_x^{\imp}$.}
To analyze PPSZ and our variant ImpatientPPSZ, we need to talk about the point in time where the algorithm processes a variable $x$, and in particular, we need to talk about the set of variables that have already been assigned a value at this point. For PPSZ, this is easy: we
define $V_x := \{y \in \vbl(F) \ | \ \pi(y) < \pi(x) \}$.
For ImpatientPPSZ, it's a bit more complicated:
imagine we run ImpatientPPSZ but feed it the ``correct'' values
in every assignment; that is, whenever a color 
$c$ is chosen, make sure that $c=d$ (we 
manipulate this random source to always choose the correct
color); pause the algorithm in the iteration when
variable $x$ is being processed, just after line \ref{line-after-while}, and
look at the partial assignment $\alpha$ built so far.
We set 
$V_x^{\imp} := \vbl(\alpha)  \setminus \{x\}$.
We remove $x$ for purely technical reasons;
if $x$ happens to be already set at that time, then line \ref{normalchoosevalue} and \ref{normalAssigned} 
will be skipped by the algorithm anyway.

\begin{observation}
    If line \ref{normalchoosevalue} is executed
    then $c$ is chosen uniformly at random from the 
    set $\plaus(x, F^{[V_x^{\rm imp} \mapsto d]})$.
\end{observation}
We define the following indicator variables:
\begin{align*}
    A_{x,c}
 & :=
 \begin{cases}
    1 & \textnormal{if } c \in \plaus(x, F^{[V_x \mapsto d]}, D) \\
    0 & \textnormal{else.}
 \end{cases} \\
    A^{\rm imp}_{x,c}
 & :=
 \begin{cases}
    1 & \textnormal{if } c \in \plaus(x, F^{[V_x^{\rm imp} \mapsto d]}, D) \\
    0 & \textnormal{else.}
 \end{cases}
\end{align*}
   and $A_x := \sum_{c} A_{x,c}$ and $A_x^{\rm imp} := \sum_c A_{x,c}^{\rm imp}$. 
   These are random variables in our random placement $\pi$. Note that $A_{x,d} = A_{x,d}^{\rm imp} = 1$ because color $d$ is always plausible;
   also, $A_{x,c}^{\rm imp} \leq A_{x,c}$ simply because
   $V_{x} \subseteq V_{x}^{\rm imp}$, i.e., ImpatientPPSZ
   has  at least as much information as PPSZ.
   \begin{lemma}\cite{hertli2016ppsz}
   For a fixed permutation $\pi$, 
       $\Pr[\textnormal{PPSZ}(F,\pi) \textnormal{ finds } \alpha^*] = \prod_{x} \frac{1}{A_x(\pi)}$. 
    For a random permutation, 
    $
        \Pr_{\pi}[\textnormal{PPSZ}(F,\pi) \textnormal{ finds } \alpha^*] \geq
        2^{- \sum_{x} \E_{\pi}[\log_2 A_x(\pi)]}
    $.
    \end{lemma}
    The second statement follows from the first
    by Jensen's inequality. 
  To obtain a similar formula for
  ImpatientPPSZ, we need to take into account that
  a variable $x$ might be assigned in
  line~\ref{impAssigned} or in line~\ref{normalAssigned}.
   \begin{lemma}\label{lemma-jensen-impatient}
   For a fixed permutation $\pi$, 
   $
       \Pr[\textnormal{PPSZ}(F,\pi) \textnormal{ finds } \alpha^*] \geq \prod_{x} \frac{1}{\max\left(1 + \eligible_x,
       A^{\rm imp}_x(\pi)\right)}.
   $
       For a random permutation, the probability 
       that ImpatientPPSZ succeeds is at least 
        \begin{align*}
        \Pr_\pi[\textnormal{PPSZ}(F,\pi) \textnormal{ finds } \alpha^*] \geq
        2^{-E_{\pi}\left[\sum_x\log_2\left(\max\left(1 + \eligible_x ,A^{\rm imp}_x(\pi)\right)\right) \right]}.
        \end{align*}
   \end{lemma}
   \begin{proof}
   If $\eligible_x = 0$ then 
   $x$ will be assigned in line 
   \ref{normalAssigned} and thus its value will 
   be correct with probability 
   $1 / A_x^{\imp}(\pi)$, conditioned
   on all prior assignments being correct.
   If $\eligible_x = 1$ then either it is assigned
   in line \ref{impAssigned}, and is correct
    with probability $1/2$; or it is still
    assigned regularly in 
    line~\ref{normalAssigned}, and is correct
    with probability $1 / A_x^{\imp}(\pi)$.
    This proves the first inequality. The second
    inequality in the lemma follows from the first
    by Jensen's inequality.
   \end{proof}

\subsection{Independence between colors}

The crucial quantity in the analysis of ImpatientPPSZ is the random variable $A_x^{\imp} =  \sum_{c} A_{x,c}^{\imp}$. 
The next lemma states that we can focus on analyzing the indicator variables 
$A_{x,c}^{\imp}$ individually; that is, if we condition on $\pi(x) = p$, then 
the $d$ indicator variables are independent
in the worst case. More formally:

\begin{lemma}[Independence between colors]\label{lemma-independence-between-colors}
  Let $\pi: V \rightarrow [0,1]$ be 
  uniformly random and set $p := \pi(x)$.
  We sample $d$ random variables
  $\tilde{A}_{x,c}^{\imp} \in \{0,1\}$, $c=1,\dots,d$ by setting each
  $\tilde{A}_{x,c}^{\imp}$ to $1$ with 
  probability
  $\Pr[A_{x,c}^{\imp} = 1 \ | \ \pi(x) = p]$,
  independently.
  Set $\tilde{A}_{x}^{\imp} := \sum_{c} \tilde{A}_{x,c}^{\imp}$.
  Then
  \begin{align}
  \label{ineq-independence-between-colors}
      \E_{\pi} \left[\log_2\left(\max\left(1 + \eligible_x ,A^{\imp}_x(\pi)\right)\right)\right]
      \leq
      \E_{\pi} \left[\log_2\left(\max\left(1 + \eligible_x ,\tilde{A}^{\rm imp}_x(\pi)\right)\right)\right]
  \end{align}
\end{lemma}

\textbf{Proof idea.} We would like to prove
this along the lines of Lemma 
3.5 of~\cite{hertli2016ppsz}. The additional
problem here is that although the function
$f: t \mapsto \log(t)$ is concave, 
the function $g: t \mapsto 
\log (\max(2, t))$ isn't. This is 
why, if $\pi(x) < \theta$, we set 
$\eligible_x$ to $1$ with probability $c$
and to $0$ with probability $1-c$. The convex 
combination $c \cdot f + (1-c) \cdot g$ is 
concave\footnote{The attentive reader might
notice: it's not concave; however,
if we change the definition of ``$\log$''
in the definition of $f$ and $g$ from
the usual $\log$ to ``$\log$ on $\mathbb{N}$
and linear between integers, then it is concave.}
and the proof goes through just 
as for Lemma 3.5 in~\cite{hertli2016ppsz}.
See Lemma~\ref{lemma-independence-between-colors-appendix}
in the appendix for a complete proof.\\

The upshot is that it is sufficient
to bound 
$\Pr[A_{x,c}^{\imp} = 1 \ | \ \pi(x) = p]$
from above, for each variable $x$ and color $c$, individually.

\subsection{Critical Clause Trees and Brief
Analysis of PPSZ}


In this section, we define  critical clause trees and review some results from \cite{hertli2016ppsz}. 
Let $x \in \vbl(F)$ and $c \in \{1,\dots,d-1\}$.  The {\em critical clause tree $T^h_{x,c}$ of height $h$} has two types of nodes: a node $u$ on an even level (which includes the root at level 0) is a {\em clause node}, has a 
{\em clause label} $\clauselabel(u)$
and an {\em assignment label}
$\beta_u$; it has at most $k-1$ children. 
A node $v$ on an odd level
is a {\em variable nodes} and has a {\em variable label}
$\varlabel(u)$; it has exactly $d-1$ children.
An edge $(v,w)$ from a variable node $v$ to a clause node
$w$ has an edge color $EC(e) \in [d-1]$.
The critical clause tree $T^h_{x,c}$ is constructed
as in algorithm \ref{Critical Clause Tree}.
\begin{algorithm}[htb] 
	\caption{BuildCCT($F,x,c,h$)}
	\label{Critical Clause Tree}
	\begin{algorithmic}[1] 
		\State Create a root node and set $\beta_{\root} := \alpha[x=c]$
		\While{$\exists$ clause node $u$ of height less than $h-1$ without a clause label}
		    \State Find a clause $C$ which is not satisfied by $\beta_u$
		    \State Set $\clauselabel(u) := C$
		    \For{each literal $(y \neq d)\in C$}
		        \State Create a new child $v$ of $u$
		        \State $\varlabel(v) := y$
		        \For{$i \in [d-1]$}
		        \State Create a new child $w$ of $v$
		        \State Set $\beta_w := \beta_v[y = i]$
		        \State Set $EC(v,w) = i$
		        \EndFor
		    \EndFor
		\EndWhile
		\State remove clause nodes at height $h+1$
		\State \textbf{return} $T^h_{x,c}$
	\end{algorithmic}
\end{algorithm}

Let us assume $h$ is always odd, so the lowest layer
of $T^h_{x,c}$ consists of variable nodes. $T^h_{x,c}$
has two types of leaves: 
those variable nodes at height $h$; we call them
{\em safe leaves}; and clause
nodes whose clause label does not contain any literal
of the form $(y \ne d)$; we call them {\em unsafe leaves}.

\begin{proposition}\cite{hertli2016ppsz}\label{CCTProperty}
\begin{enumerate}
    \item Suppose $v$ is a clause node in $T^h_{x,c}$ with clause label $C$ and $(y \neq i), i\in [d]$ is a literal in $C$. Then if $i=d$, $v$ has a child whose variable label is $y$. If $i<d$, $v$ has an ancestor node whose variable label is $y$.
    \item No variable appears more than once 
    as variable label on a path from
     root to a leaf.
\end{enumerate}
\end{proposition}

\begin{definition}[labeled tree]
A labeled tree is a possibly infinite tree such that: (1) every node is either a variable node or a clause node; (2) a variable node $u$ has a label
$\varlabel(u) \in \mathbb{L}$ in some 
label space $\mathbb{L} \supseteq V$; (3) they alternate, i.e., if a variable node has children, they are all clause nodes, and vice versa;
(4) its degree is bounded:
there is some $\Delta \in \mathbb{N}$
such that every node has at most $\Delta$
children.
A leaf in a labeled tree is a safe leaf if it is a variable node; Otherwise, it is an unsafe leaf.
\end{definition}
Note that each subtree of a critical clause tree is a labeled tree.
A {\em safe path} in a labeled tree
is a path that starts at the root
and is either infinite or 
ends at a safe leaf.
\begin{definition}[$\cut_p$ and $\cut$]
Let $T$ be a labeled tree. 
The event $\cut_p(T)$ is an event in the probability
space of all placements $\pi: \mathbb{L} \rightarrow [0,1]$ that happens if 
every safe path in $T$ contains a 
node $v$ with $\pi(\varlabel(v)) < p$.
\end{definition}

Suppose $T$ is a labeled tree, and let 
$T_1,\dots, T_l$ be the subtrees rooted at the $l$ children
of the root of $T$. Note that the $T_i$ are themselves 
labeled trees. If the root of $T$ is a clause node then
$\cut_p(T) = \bigwedge_{i=1}^l \cut_p(T_i)$.
If it is a variable node, let $y := \varlabel(\root(T))$,
and observe that 
$\cut_p(T) = [\pi(y) < p]$ if $\root(T)$ itself is a safe leaf
(i.e., if $l = 0$)
and 
$\cut_p(T)  = [\pi(y) < p] \vee \bigwedge_{i=1}^{l} \cut_p(T_i)$
else .\\

Next, we connect the notion of cuts to our 
notion of being a plausible color. For this, 
set $L := (d-1)(k-1)$ and observe 
 that $T^h_{x,c}$ has at most $L^i$ clause nodes
 at depth $2i$.
 Choose $\tilde{h}$ to be the largest integer for which 
  $1 + L + L^2 + \dots + L^{\tilde{h}} \leq D$ (recall
  $D$, our strength parameter in the definition of $D$-implication),
  and set $h := 2\, \tilde{h} + 1$.
 Then $T^h_{x,c}$ has at most $D$ clause nodes 
 and $h$ is also a slowly growing function in $n$.

\begin{lemma}[\cite{hertli2016ppsz}]
\label{lemma-LocalReasoning}
If  $\cut(T^h_{x,c})$ happens then 
$A_{x,c} = 0$.
\end{lemma}

Recall the infinite trees $T^\infty$ and
$T_1,\dots,T_{d-1}$  and the
indicator variables $J_1,\dots,J_{d-1}$ defined
above, just before~(\ref{eq-definition-Sdk}), 
and observe that $J_c=1$ iff $\cut_p(T_c)$ does {\em not} happen.
Let $T_\infty$ be the subtree of $T^\infty$ rooted at the first child of the root.
Define $Q(p) := \Pr[\cut_p(T^{\infty})]$
and $R(p) := \Pr[\cut_p(T_\infty)]$.
The next proposition is 
from~\cite{hertli2016ppsz}, adapted 
for our purposes.
\begin{proposition}[\cite{hertli2016ppsz}]
\label{ExtinProbPPSZ}
   Set $L = (k-1)(d-1)$. If $p \geq 1 - \frac{1}{L}$
   then $Q(p) = R(p) = 1$; otherwise, $Q(p)$ and $R(p)$ are the unique 
   roots in $[0,1]$ of the equations
   $
       Q =   \left( p + (1-p) Q^{d-1} \right)^{k-1} 
   $ and 
   $       R = p + (1-p) R^L$, respectively. Furthermore, 
   $Q(p) = R(p)^{k-1}$.
 \end{proposition}

As our height parameter $h$ grows (roughly
logarithmic with  our strength parameter $D$),
the critical clause trees $T^h_{x,c}$ will 
look more and more like $T^{\infty}$,
and thus the cut probability will converge 
to $Q(p)$. Formally,  let $\error(d,k,h,p)$ 
 and $\error(d,k,h)$ stand for any functions that 
 converge to $0$ as $h \rightarrow \infty$.

\begin{proposition}[Lemma 3.6 in~\cite{hertli2016ppsz}]\label{lemma-cut-prob-Q}
   $\Pr[\cut_p(T^h_{x,c})] \geq \Pr[\cut_p(T_c)] - \error(d,k,p,h)$.
\end{proposition} 
 To summarize: 
 conditioned on $\pi(x) = p$, the sum
 $A_x = A_{x,1} + \cdots + A_{x,d}$ has the 
 worst behavior if all $A_{x,c}$ are independent
 (Lemma~\ref{lemma-independence-between-colors});
 furthermore, $A_{x,c} \leq J_c$ except
 with probability $\error(d,k,p,h)$, for all $c \leq d-1$,
 and therefore:
 \begin{lemma}\label{AtoS}\cite{hertli2016ppsz}
     $\E_{\pi}[ \log_2 (A_x)] \leq
      \E[\log_2(J_1 + \cdots + J_{d-1} + 1)] + 
      \error(d,k,h) 
      = S_{d,k} + \error(d,k,h)$.
 \end{lemma}

\section{Analysis of  ImpatientPPSZ} 

Just as \cite{hertli2016ppsz} analyzes PPSZ 
by studying the random variables
$A_{x,c}$, we have to study $A_{x,c}^{\imp}$.
We can always resort to the ``old'' analysis via $A_{x,c}^{\imp} \leq A_{x,c}$. However,
the whole point of this work is to show
that this inequality is often strict.
To understand how and when this might happen, we discuss  an example for $d=3$.

\begin{center}
    \includegraphics[width=0.5\textwidth]{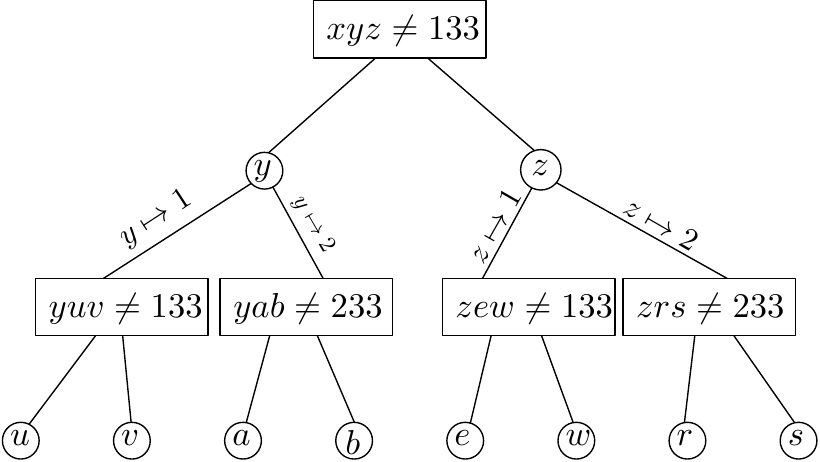}
\end{center}

This is $T^3_{x,1}$, the critical clause tree
for $x$ and $1$ built up to height $3$.
The formula $F$ in question contains the constraints
shown as clause labels, but of course contains many
more constraints. Suppose 
that $u,v,a,b,z$ come before $x$ in $\pi$,
and $e,w,r,s,y$ come later. In the normal PPSZ,
we have already set $u,v,a,c,z \mapsto 3$ when considering
$x$, and thus the clauses of $F$
will have shrunk:
\begin{itemize}
    \item $(yuv \ne 133)$ shrinks to $(y \ne 1)$;
    \item $(yab \ne 133)$ shrinks to $(y \ne 2)$;
    \item $(zew \ne 133)$ and $(zrs) \ne 233$ don't shrink
    but disappear: they are satisfied by $z \mapsto 3$;
    \item $(xyz \ne 133)$ shrinks to $(xy \ne 13)$.
\end{itemize}
Together, the three shrunk clauses
$(y \ne 1)$, $(y \ne 2)$, and $(xy \ne 13)$ imply $(x \ne 1)$; since $D \geq 3$  this means 
that $x=1$ can be ruled out, i.e., $A_{x,1} = 0$.
Next, suppose $\pi$, viewed as a placement
$\pi: V \rightarrow [0,1]$, looks like this:

\begin{center}
    \includegraphics[width=0.7\textwidth]{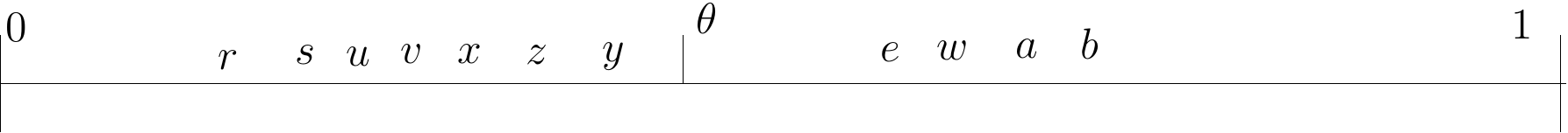}
\end{center}
 and assume for simplicity that
 all variables $l$ 
with $\pi(l) < \theta$ in $\vbl(F)$ are eligible for impatient
assignment (i.e., have $\eligible_l = 1$).
Note that $\cut(T^3_{x,1})$ does 
not happen. Namely, the path
from root to $c$ contains two variable labels, $y$ and $b$,
and $\pi(y), \pi(b) \geq \pi(x)$. Analogously, the 
alternative assignment $\alpha^*[x\mapsto 1, y\mapsto 2,b\mapsto 2]$ 
satisfies all clauses in the figure above, and thus 
the algorithm cannot infer $x \ne 1$
from those clauses alone, and $A_{x,1} = 1$. Observe
now what happens in ImpatientPPSZ: 
\begin{itemize}
    \item $r,s,u,v \mapsto 3$ before $x$ is even considered;
    \item $(yuv \ne 133)$ shrinks to $(y \ne 1)$,
    and thus $\plaus(y,F^{[\alpha]})$ shrinks to $\{2,3\}$;
    \item $y$ is assigned a value in line 6;
    \item the analogous thing happens to $z$;
    \item $r,s,u,v \in V_x$, and $r,s,u,v,y,z \in V^{\imp}_x$;
    \item $(xyz \ne 133)$ shrinks to $(x \ne 1)$ and 
    thus $A^{\imp}_{x,1}=0$.
\end{itemize}

We can now try to work out a formula for the probability that $x=c$ is ruled out in this manner; however, our 
above example and analysis contains two silent assumptions that cannot be taken for granted in general:
\begin{enumerate}
    \item All variable labels in $T^3_{x,c}$ are distinct.
    \item All clause labels of $T^{3}_{x,c}$ are 
    critical clauses, i.e., $k-1$ of its literals
    are of the form $y \ne d$.
\end{enumerate}
The original PPSZ paper~\cite{paturi2005improved}
addresses Point 1 by
using the FKG inequality to show that having
multiple labels can never hurt us. But now we are talking
about a more complicated event; it is not clear whether
an FKG-like result applies. Point 2 is more troublesome. Consider the alternative
scenario that 
$T^3_{x,c}$ looks like this:

\begin{center}
    \includegraphics[width=0.5\textwidth]{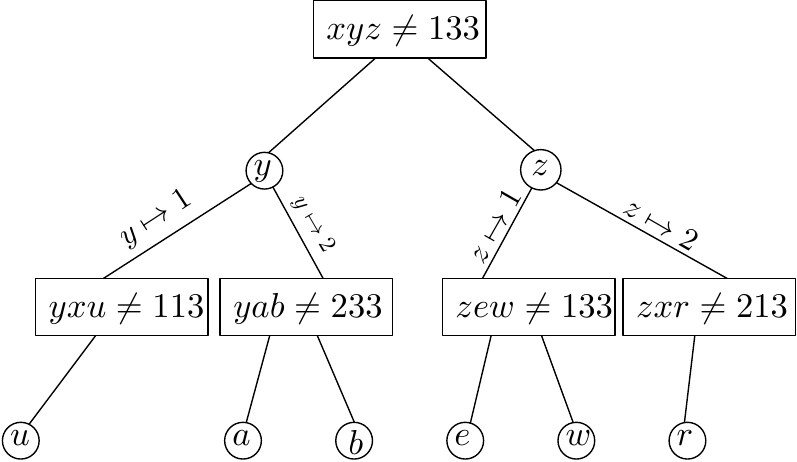}
\end{center}
and consider the same $\pi$ as above:
$r,s,u,v,x,z,y, \theta, e,w,a,b$. After 
setting $r,s,u,v\mapsto 3$, the shrunk clauses
are $(yx \ne 11)$, $(yab \ne 233)$,
$(zew \ne 133)$, and $(zx \ne 21)$. Neither
for $y$ nor for $z$ can we rule out any 
color, and therefore our impatient mechanism
will not kick in. We will have
$V_x = V_x^{\imp} = \{r,s,u,v\}$. In other 
words, non-critical clauses seem
useless for ImpatientPPSZ. But looking
at the above example tree,
we see what comes to the rescue: 
the right-most clause node 
is missing a child; it 
has at most $k-2$ children instead of $k-1$.
This alone will be enough to improve 
our success probability by a bit.
It is time for some formal definitions.

\begin{definition}[Privileged variables]
A variable $x$ is {\em privileged} if there is 
some color $c \in \{1,\dots, d-1\}$ such that
\begin{enumerate}
    \item $T^h_{x,c}$ 
    has fewer than $(k-1)^2 (d-1)$ variable nodes 
    at level 3 {\em or}
    \item  
    $T^3_{x,c}$ has two variable nodes $u$ and $w$
    with $\varlabel(u) = \varlabel(w)$.
\end{enumerate}
\end{definition}

\newcommand{\privileged}{\textnormal{privileged}}
\begin{proposition}
\label{proposition-privileged-PPSZ}
There is an $\epsilon_{\privileged} > 0$, depending
only on $d$ and $k$, such that
\begin{align*}
    \costspelledout \leq S_{d,k} - 
    \epsilon_{\privileged} + \error(d,k,h)\  ,
\end{align*}
for every privileged variable $x$ in $F$.
\end{proposition}
See Proposition~\ref{proposition-privileged-PPSZ-appendix} in the appendix for a proof. 
\begin{corollary}
\label{corollary-privileged-ImpatientPPSZ}
 $\costimpspelledout \leq S_{d,k} - 
    \epsilon_{\privileged} + c \theta
    + \error(d,k,h)$
\end{corollary}
\begin{proof}
 Since
    $\max(a,b) \leq a \cdot b$ when $a,b \geq 1$,
    we get
    \begin{align*}
       \costimpspelledout & \leq
        \E_{\pi} [ \log_2(1 + \eligible_x)]
        + 
        \E_{\pi} [ \log_2( A^{\imp}_x) ] \ . 
    \end{align*}
    The first term equals
    $\Pr[\eligible_x = 1] = c \theta$;
    the second is at most 
    $\costspelledout$, which by
    Proposition~\ref{proposition-privileged-PPSZ}
    is at most $S_{d,k} - \epsilon_{\privileged} + 
    \error(d,k,h)$. This concludes the proof.
\end{proof}

\begin{lemma}\label{lemma-improvement-non-privileged}
  There is a constant $\epsilon > 0$, 
  depending only on $d$ and $k$, such that
  \begin{align*}
   \costimpspelledout \leq S_{d,k} - 0.1699 \, \left( 
    \frac{c}{L+1} \theta^{L+1} + O\left( \theta^{L+2} \right)
    \right)
    + \error(d,k,h) \ . 
\end{align*}
for all non-privileged variables $x$.
The constant factor hidden in the $O(\cdot)$ 
depends only on $d$ and $k$.
\end{lemma}

By choosing $\theta$ sufficiently small,
we can make sure that the bounds in
Lemma~\ref{lemma-improvement-non-privileged}
and Corollary~\ref{corollary-privileged-ImpatientPPSZ}
are both at most $S_{d,k} - \epsilon_{d,k} + 
\error(d,k,h)$, for some $\epsilon_{d,k}$ depending
only on $d$ and $k$. Together with Lemma~\ref{lemma-jensen-impatient}, this
proves Theorem~\ref{theorem-main}.
    
\begin{proof}[Proof of Lemma~\ref{lemma-improvement-non-privileged}]
    For a color $1 \leq c \leq d-1$, 
    fix the critical clause tree $T^h_{x,c}$ 
    and let us introduce a bit of notation.
    The root of $T^h_{x,c}$ has a label
    \begin{align*}
        C_{\root} = (x \ne c \vee y_1 \ne d \dots\vee y_{k-1} \ne d) \ .
    \end{align*}
    It has $k-1$ children $v_1,\dots,v_{k-1}$,
    whose respective variable labels are 
    $y_1,\dots,y_{k-1}$. Let $T_i$ denote
    the subtree of $T^h_{x,c}$ rooted at $v_i$.
    Each $y_i$ in turn
    has $d-1$ children; each such level-2 node $v$
    has a clause label $C_v$; note that 
    $C_v$ is a {\em critical clause}, i.e.,
    $k-1$ of its literals are of the form
    $(z \ne d)$, since otherwise it would have fewer than
    $k-1$ children, and $T^h_{x,c}$ would have fewer 
    than $(k-1)^2 (d-1)$ nodes at level 3; in other
    words, $x$ would be privileged.
    
    We need to define an event $\impcut_p(T^h_{x,c})$ which,
    analogous to $\cut_p(T^h_{x,c})$, describes the event
    $A^{\imp}_{x,c} = 0$ in terms of $T^h_{x,c}$ only.
    Going for a full such characterization is possible
    but messy, and it is not clear what the
    worst-case structure of such $T^h_{x,c}$ will be; this is 
    the reason why we, when considering
    our impatient assignment mechanism, will look only 
    up to depth $3$ in $T^h_{x,c}$.
    For each node $w$ of $T^h_{x,c}$ at level $1$, $2$, or $3$, 
    we define 
    event $\localimpcut_p(v)$ as follows:\\
    \begin{enumerate}
        \item If $v$ is at level 3 of $T^h_{x,c}$ then
        $\localimpcut_p(v)$ happens if 
        $\pi(\varlabel(v)) < p$.
        \item If $v$ is at level $2$ of $T^h_{x,c}$
        then $\localimpcut_p(v)$ happens if 
        $\localimpcut_p(w)$ happens for the $k-1$
        children $w$ of $v$ (recall that $\clauselabel(v)$
        is a critical clause and therefore $v$ has exactly
        $k-1$ children);
        \item If $v$ is at level $1$, set $y := \varlabel(v)$; 
        $\localimpcut_p(v)$ happens if 
        \begin{enumerate}
            \item $\pi(y) < p$ {\em or}
            \label{impcut-y-regular}
            \item $\eligible_{y}= 1$ {\em and}
            $\localimpcut_p(v)$ happens for at least
        $d-2$ of the $d-1$ children of $v$.
        \label{impcut-y-early}
        \end{enumerate}
    \end{enumerate}
    Finally, we define
    \begin{align}
    \label{eq-impcut}
        \impcut_p(T^h_{x,c}) := \bigwedge_{i=1}^{k-1}
        \left(
         \cut_p(T_i)
         \vee
         \localimpcut_p(v_i)
         \right)
    \end{align}
    The next lemma is the ``impatient analog''
    of Lemma~\ref{lemma-LocalReasoning}.
    \begin{lemma}
    \label{lemma-impatient-cut}
        Let $p = \pi(x)$.
        If $\impcut_{p}(T^h_{x,c})$ happens then
        $A^{\imp}_{x,c} = 0$.
    \end{lemma}
    The proof is very similar to that
    of Lemma~\ref{lemma-LocalReasoning}, just 
    taking into account the impatient assignment
    mechanism. We restate and prove
    it as 
    Lemma~\ref{lemma-impatient-cut-appendix}
    in the appendix.
Next, we prove a lower bound on $\Pr[\impcut_p(T^h_{x,c})]$.
For $q \in [0,1]$ and $l \in \mathbb{N}$, define
\begin{align}
\label{eq-abamo}
    \abamo(q, l) := q^l + l (1-q) q^{l-1} \ .
\end{align}
The name $\abamo$ is the acronym of ``all 
but at most one'' and is indeed the probability  
that, among $l$ independent events of probability $q$
each, all or all but one happen.
Recall the definition of $Q(p) := 
\Pr[\cut_p(T^{\infty})]$ just before 
Proposition~\ref{ExtinProbPPSZ}.

\begin{lemma}
\label{lemma-cut-prob-impatient}
  If $p < \theta$ then $  \Pr[\impcut_p(T^h_{x,c}) \ | \ \pi(x) = p]$ is at least
  \begin{align*}
      \left( p + c (\theta -p) \abamo( p^{k-1}, d-1)
       + (1 - p - c(\theta - p)) Q(p)^{d-1}
      \right)^{k-1} - \error(d,k,h) \ . 
  \end{align*}
  If $p \geq \theta$ then it is at least $Q(p)-\error(d,k,h)$.
\end{lemma}
\textbf{Proof sketch.} For each subtree $T_i$ 
of $T^h_{x,c}$, either $\cut_p(T_i)$
or $\localimpcut_p(v_i)$ must happen.
Now this happens if either (1) $\pi(y) < p$,
which explains the first term of the sum
in the parentheses; (2) $\pi(y) \geq p$
and $\eligible_{y_i}=1$ and $\localimpcut_p(v_i)$,
which is the second term; or (3) 
$\pi(y) \geq p$ and $\eligible_{y_i} = 0$
and $\cut_p(T_i)$, which is the third term.
See Lemma~\ref{lemma-cut-prob-impatient-appendix}
for a complete proof.\\

Let us summarize our reasoning so far. 
Define  an ensemble 
$J^{\imp}_1,\dots, J^{\imp}_{d-1}$
of random variables in $\{0,1\}$
as follows: set $p := \pi(x)$;
then independently set each $J^{\imp}_c$ to $0$ with probability
$W^{k-1}$ and $1$ with probability $1 - W^{k-1}$, where 
\begin{align*}
    W = W(p) := 
    \begin{cases}
    p + c (\theta -p) \abamo( p^{k-1}, d-1)
       + (1 - p - c(\theta - p)) Q(p)^{d-1} & 
       \textnormal{ if $p < \theta$} \\
       R(p) & \textnormal{ else.}
       \end{cases}
\end{align*}
One checks that $W(p)$ is continuous at $p = \theta$
since $R(p) = p + (1-p) R(p)^{(k-1)(d-1)} = 
p+  (1-p) Q(p)^{d-1}$.
Set $J^{\imp} := J^{\imp}_1 + \cdots + J^{\imp}_{d-1} + 1$.
We have shown so far that
\begin{align}
    \label{ineq-exp-A-J}
    \E\left[\log_2\max(1 + \eligible_x, A^{\imp}_{x,c} ) \right] & \leq
    \E\left[\log_2 \max(1 + \eligible_x, J^{\imp}) \right] + 
    \error(d,k,h)\nonumber \\
    & = 
    \Pr[J^{\imp} = 1 \wedge \eligible_x] 
    + 
    \E\left[\log_2  (J^{\imp}) \right]
    + \error(d,h,k) \ . 
\end{align}
\begin{proposition}
\label{proposition-loss}
   $\Pr[ J^{\imp} = 1 \wedge \eligible_x]  \leq 
   \frac{c}{L+1} \theta^{L+1}
   + O\left( \theta^{L+2} \right)$.
\end{proposition}

\begin{proposition}
\label{proposition-gain}
    $\E\left[\log_2  (J^{\imp})\right]- S_{d,k} \leq
    (d-1) \log_2 (1 - 1/d) \cdot 
    \left(
    \frac{c}{L+1} \theta^{L+1} + O\left( \theta^{L+2} \right)
    \right)
    $.
\end{proposition}
We prove the two propositions in 
Section~\ref{section-appendix-loss-gain} in the appendix.
Together with (\ref{ineq-exp-A-J}), they
imply that 
$\E\left[\log_2  \max(1 + \eligible_x, A^{\imp}_{x,c} ) \right]
- S_{d,k}$ is at most 
\begin{align*}
& 
    \left( 
    \frac{c}{L+1} \theta^{L+1} + O\left( \theta^{L+2} \right)
    \right)
     \left( 1  + (d-1) \log_2 (1-1/d)  \right)
    + \error(d,k,h) \ . 
\end{align*}
The expression in the first parenthesis 
is positive for sufficiently
small $\theta$; in fact, we have to choose $\theta$ small 
enough to beat the hidden constant in the $O(\cdot)$, which
in turn depends only on $d$ and $k$. The expression
in the second parenthesis, $1 + (d-1) \log_2(1 - 1/d)$, 
is negative for all $d \geq 3$. It is
maximized for $d=3$,
where it becomes $2 - 2\, \log_2(3) < 
- 0.1699$.
Thus, we can choose $\theta$ such that 
the whole expression is at most 
$S_{d,k} - \epsilon_{d,k} + \error(d,k,h)$ for some
$\epsilon_{d,k} > 0$ depending only on $d$ and $k$.
This concludes the proof of Lemma~\ref{lemma-improvement-non-privileged}.  
\end{proof}


\section{Future Work}
\label{sec:Conclusion}
In the analysis of PPSZ, the worst case happens if all 
everything looks ``nice'': all variable 
nodes in  $T_{x,1},\dots, T_{x,d-1}$  have 
different labels; all clause labels 
are critical clauses.

In this scenario, our analysis for impatient assignment
could go deeper than level 3; we could define
a more powerful event $\impcut$ and obtain 
much better bounds on the running time.
Indeed, future work hopefully will 
identify the worst-case shape of the $T^h_{x,c}$
and allow us to analyze the full power impatient assignment.

The condition
$|\plaus(y, F^{[\alpha]})| \leq 2$ in 
Line~\ref{line-impatient-ppsz-while}
in Algorithm~\ref{ImpatientPPSZ} is arbitrary.
Why ``$\leq 2$''? Why not ``$\leq 3$''? For large
$d$, what would the optimal cut-off value be?

\section*{Acknowledgments}

Dominik Scheder wants to thank
Timon Hertli, Isabelle Hurbain, Sebastian Millius, Robin A. Moser, and May Szedl\'ak, his co-authors
of~\cite{hertli2016ppsz}.
The idea of 
impatient assignment came up when we were 
working on~\cite{hertli2016ppsz}.


\bibliographystyle{unsrt}  
\bibliography{references}  

\begin{thebibliography}{10}

\bibitem{beigel20053}
Richard Beigel and David Eppstein.
\newblock {3-coloring in time $O\left(1.3289^n\right)$}.
\newblock {\em J. Algorithms}, 54(2):168--204, 2005.

\bibitem{schoning1999probabilistic}
Uwe Sch{\"o}ning.
\newblock A probabilistic algorithm for {$k$}-{SAT} and constraint satisfaction
  problems.
\newblock In {\em Proceedings of the 40th {A}nnual {S}ymposium on {F}oundations
  of {C}omputer {S}cience}, pages 410--414. IEEE Computer Society, Los
  Alamitos, CA, 1999.

\bibitem{paturi1997satisfiability}
Ramamohan Paturi, Pavel Pudl{\'a}k, and Francis Zane.
\newblock {Satisfiability coding lemma}.
\newblock In {\em Proceedings 38th Annual Symposium on Foundations of Computer
  Science}, pages 566--574. IEEE, 1997.

\bibitem{paturi2005improved}
Ramamohan Paturi, Pavel Pudl{\'a}k, Michael~E Saks, and Francis Zane.
\newblock {An improved exponential-time algorithm for k-SAT}.
\newblock {\em Journal of the ACM (JACM)}, 52(3):337--364, 2005.

\bibitem{scheder2010ppz}
Dominik Scheder.
\newblock {PPZ for more than two truth values-an algorithm for constraint
  satisfaction problems}.
\newblock {\em arXiv preprint arXiv:1010.5717}, 2010.

\bibitem{hertli2016ppsz}
Timon Hertli, Isabelle Hurbain, Sebastian Millius, Robin~A Moser, Dominik
  Scheder, and May Szedl{\'a}k.
\newblock {The PPSZ algorithm for constraint satisfaction problems on more than
  two colors}.
\newblock In {\em International Conference on Principles and Practice of
  Constraint Programming}, pages 421--437. Springer, 2016.

\bibitem{hertli2011}
Timon Hertli.
\newblock 3-{SAT} faster and simpler---unique-{SAT} bounds for {PPSZ} hold in
  general.
\newblock In {\em 2011 {IEEE} 52nd {A}nnual {S}ymposium on {F}oundations of
  {C}omputer {S}cience---{FOCS} 2011}, pages 277--284. IEEE Computer Soc., Los
  Alamitos, CA, 2011.

\bibitem{SchederSteinberger}
Dominik Scheder and John~P. Steinberger.
\newblock {PPSZ for General k-SAT - making Hertli's analysis simpler and 3-SAT
  faster}.
\newblock In Ryan O'Donnell, editor, {\em 32nd Computational Complexity
  Conference, {CCC} 2017, July 6-9, 2017, Riga, Latvia}, volume~79 of {\em
  LIPIcs}, pages 9:1--9:15. Schloss Dagstuhl - Leibniz-Zentrum fuer Informatik,
  2017.

\bibitem{HKZZ}
Thomas~Dueholm Hansen, Haim Kaplan, Or~Zamir, and Uri Zwick.
\newblock {Faster $k$-SAT algorithms using biased-PPSZ}.
\newblock In Moses Charikar and Edith Cohen, editors, {\em Proceedings of the
  51st Annual {ACM} {SIGACT} Symposium on Theory of Computing, {STOC} 2019,
  Phoenix, AZ, USA, June 23-26, 2019}, pages 578--589. {ACM}, 2019.

\bibitem{scheder2021eccc}
Dominik Scheder.
\newblock {PPSZ} is better than you think.
\newblock {\em Electron. Colloquium Comput. Complex.}, 28:69, 2021.

\end{thebibliography}

\appendix
\section{Independence between colors}

\begin{lemma}[Lemma \ref{lemma-independence-between-colors}, restated]\label{lemma-independence-between-colors-appendix}
  Let $\pi: V \rightarrow [0,1]$ be 
  uniformly random and set $p := \pi(x)$.
  We sample $d$ random variables
  $\tilde{A}_{x,c}^{\imp} \in \{0,1\}$, $c=1,\dots,d$ by setting each
  $\tilde{A}_{x,c}^{\imp}$ to $1$ with 
  probability
  $\Pr[A_{x,c}^{\imp} = 1 \ | \ \pi(x) = p]$,
  independently.
  Set $\tilde{A}_{x}^{\imp} := \sum_{c} \tilde{A}_{x,c}^{\imp}$.
  Then
  \begin{align}
  \label{ineq-independence-between-colors-appendix}
      \E_{\pi} \left[\log_2\left(\max\left(1 + \eligible_x ,A^{\imp}_x(\pi)\right)\right)\right]
      \leq
      \E_{\pi} \left[\log_2\left(\max\left(1 + \eligible_x ,\tilde{A}^{\rm imp}_x(\pi)\right)\right)\right]
  \end{align}
\end{lemma}

\begin{proof}
   We prove 
   (\ref{ineq-independence-between-colors-appendix})
   conditioned on $\pi(x) = p$.
   Let $\mathbf{Z} \in \{0,1\}^{V \setminus \{x\}}$
   be defined by $Z_y := [\pi(y) \geq p]$.
   Note that each $Z_y$ is $1$ with probability
   $1-p$, independently. 
   Next, observe that each $A_{x,c}^{\imp}$
   is a monotone increasing Boolean function
   $f_c(Z)$: moving some $\pi(y)$ above
   $p$ can only increase $A_{x,c}^{\imp}$. 
   Let $\mathbf{Z}^{(1)}, \dots, \mathbf{Z}^{(d)}$
   be $d$ independent copies of $\mathbf{Z}$;
   that is, each has the same distribution
   as $\mathbf{Z}$ but they are independent. 
   Conditioned on $\pi(x) = p$, we have 
   $(f_1(\mathbf{Z}), \dots, f_d(\mathbf{Z}))
   \sim (A_{x,1}^{\imp}, \dots, A_{x,d}^{\imp})$
   and 
   $(f_1(\mathbf{Z}^{(1)}), 
   \dots, f_d(\mathbf{Z}^{(d)})) \sim
   (\tilde{A}_{x,1}^{\imp},\dots,\tilde{A}_{x,d}^{\imp}) $,
   where $A \sim B$ means that the random variables $A$ and $B$ have the same distribution.
   
   Now if $p > \theta$ and therefore $\eligible_x = 0$,
   then the function $\log_2 (\max(1 + \eligible_x, \cdot))$ in (\ref{ineq-independence-between-colors-appendix}) becomes
   $\log_2(\cdot)$ and we can directly
   apply the Concave Correlation Lemma (Lemma A.1
   of the full version of~\cite{hertli2016ppsz}).
   
   If $\eligible_x = 1$, the trouble
   is that the function
   $t \mapsto \log_2(\max (2, t))$ is not concave anymore. However, note that
   if $\pi(x) < \theta$, we set
   $\eligible_x$ to $1$ with probability 
   $c$ and $0$ with probability $1-c$. Conditioned on $\pi(x)=p$, the randomness in (\ref{ineq-independence-between-colors-appendix})
   comes from two sources: 
   (1) the choice of $\eligible_x$;
   (2) the randomness in $\mathbf{Z}$
   (or $\mathbf{Z}^{(1)},\dots,\mathbf{Z}^{(d)}$ for the right-hand side).
   We can break down both sides of
  (\ref{ineq-independence-between-colors-appendix}) as follows:
   \begin{align}
   \label{ineq-with-c}
       \E_{\mathbf{Z},\eligible_x}
       [ \log_2(\max(1 + \eligible_x, 
       A^{\imp}_x))] =
    \E_{\mathbf{Z}} 
    \left[c\log_2(\max(2,A^{\imp}_x))
    + (1-c) \log_2(A^{\imp}_x)\right] \ ,
   \end{align}
   where $c = 2 - \log_2(3)$.
   Now  the function
   $t \mapsto c \log_2(\max(2,t)) 
   + (1-c) \log_2(t)$ is still not 
   concave. 
   However, note that the arguments
   of $\log_2(t)$ in
   (\ref{ineq-independence-between-colors-appendix})
   and (\ref{ineq-with-c}) are integers;
   define $g(t)$ to be the function that
   equals $\log_2(t)$ if $t$ is an integer,
   and is linear between integers. 
   Now $g$ is concave and 
   $t \mapsto c g(\max(2,t)) 
   + (1-c) g(t)$ is concave, too.
   In fact, this function is linear 
   on $[1,3]$ and agrees with $g$
   for $t \geq 3$.
   Now the lemma again follows
   by the Concave Correlation Lemma
   (Lemma A.1 of~\cite{hertli2016ppsz}).
\end{proof}

\section{PPSZ for privileged variables}
\begin{proposition}[Proposition~\ref{proposition-privileged-PPSZ}, restated]
\label{proposition-privileged-PPSZ-appendix}
Suppose $x\in \vbl(F)$ is a priviledged variable. Then there is an $\epsilon_{\privileged} > 0$, depending
only on $d$ and $k$, such that
\begin{align*}
    \costspelledout \leq S_{d,k} - 
    \epsilon_{\privileged} + \error(d,k,h)\  ,
\end{align*}
for every privileged variable $x$ in $F$.
\end{proposition}
\begin{proof}
This proof is similar in spirit and also
technical details to the proof of Lemma 19 
in~\cite{scheder2021eccc}, except that the latter
is concerned with SAT (i.e., the case $d=2$).\\

Note that a variable $x$ can be $\privileged$
for two reasons: first, there is some color $c$
such that the critical clause tree 
$T^h_{x,c}$ has fewer than $(k-1) L$ leaves 
at level 3; in other words, some clause node
$v$ at level $2$ has fewer than $k-1$ children
(note that the nodes at level 0 and 1 have 
the ``right'' numer of children; the clause
label of $0$ is a critical clause, 
and therefore the root has always $k-1$
children; an odd-level node always has $d-1$
children). The second reason
would be that, for some color $c$,
level 1 and 3 of the critical clause
tree $T^h_{x,c}$ contain nodes $u$ and $v$
with $\varlabel(u) = \varlabel(v)$.\\

It is easy to see that the first kind of 
privilege is stronger: let $v$ be the level-2
node with fewer than $k-1$ children. 
We can add ``fictitious'' subtrees until
$v$ has $k-1$ children, and make sure 
that one of the added children shares 
its variable label with an already-existing
level-3 node. The result of this operation,
$T'_{x,c}$, exhibits a privilege of the second
kind, and $\cut_p(T^h_{x,c}) \supseteq \cut_p(T'_{x,c})$.\\

Thus, let us assume that $x$ is privileged
because $T^h_{x,c}$ contains two
nodes $v$ and $w$ with $\varlabel(v) = \varlabel(w) = z$
and the depths of $v$ and $w$ are in $\{1,3\}$.
Analogous to the proof of Proposition~\ref{lemma-cut-prob-Q} (Lemma 3.5 in~\cite{hertli2016ppsz}, we start with
iteratively
assign fresh labels to variable nodes;
as shown in~\cite{hertli2016ppsz}, this never increases $\Pr[\cut_p(T^h_{x,c})]$. 
We apply this to all
variable nodes except $v$ and $w$, and obtain
a new tree $T$. We make sure
that there are no ``missing children''
in $T$, i.e., that 
every clause has $k-1$ children; 
this can be achieved by attaching
fictitious subtrees, which does not 
increase $\Pr[\cut_p(T)]$.
Also, we will for convenience
assume that $T$ is infinite, i.e.,
has no safe leaves (and no unsafe leaves, either). This does increase $\Pr[\cut_p]$,
but by at most $\error(d,k,h)$.
In $T$ we still have 
$\varlabel(v) = \varlabel(w) = z$, but all 
other labels are distinct. Let $T'$ be the 
tree where $v$ and $w$ receive fresh
labels $z_v, z_w$. We already know that 
    $\Pr[\cut_p(T')] = Q(p)$.
It remains to show that 
$\Pr[\cut_p(T)]$ is substantially larger
than $\Pr[\cut_p(T')]$. For this, let $\mathbb{L}$
be the set of variable labels appearing in $T$ and $T'$, 
and let $\tau: \mathbb{L} \setminus \{z,z_v,z_w\} 
\rightarrow [0,1]$. We will analyze the difference
\begin{align}
\label{eq-difference-T-Tprime}
    \Pr[\cut_p(T) \ | \ \tau] - \Pr[\cut_p(T') \ | \ \tau]
\end{align}
for fixed $\tau$. Introduce the three Boolean variables
$a := [\pi(z) < p]$, $a_v := [\pi(z_v) < p]$, and
$a_w := [\pi(z_w) < p]$. Note that under $\tau$,
the event $\cut_p(T')$ reduces to $f_{\tau}(a_v, a_w)$ 
for some monotone 
Boolean function and $\cut_p(T)$ reduces
to $f_{\tau}(a, a)$, for the same function $f_\tau$.
There are only six possible such functions:
$f_\tau(a_v, a_w)$ is either $0$, $1$, $a_v$, $a_w$,
$a_v \wedge a_w$, or $a_v \vee a_w$.
If it is one of the first four, then 
$\Pr[f_\tau(a_v, a_w)] = \Pr[f_{\tau}(a,a)]$ and
(\ref{eq-difference-T-Tprime}) is $0$.
It cannot be $a_v \vee a_w$: the nodes $v$ and $w$ are
not ancestors of each other. Finally,
if $f_\tau(a_v, a_w) = a_v \wedge a_w$ then
we call $\tau$ {\em pivotal}
and observe that
(\ref{eq-difference-T-Tprime}) becomes
$p - p^2$. 

From here on, our plan is to 
lower bound the probability
that $\tau$ is pivotal.
We give a necessary and sufficient
criterion for $\tau$ to be 
pivotal.\footnote{Actually, it is sufficient for our purposes that the criterion be sufficient, and not necessary that it be necessary.}
It is best illustrated
with a figure.
\begin{center}
    \includegraphics[width=0.6\textwidth]{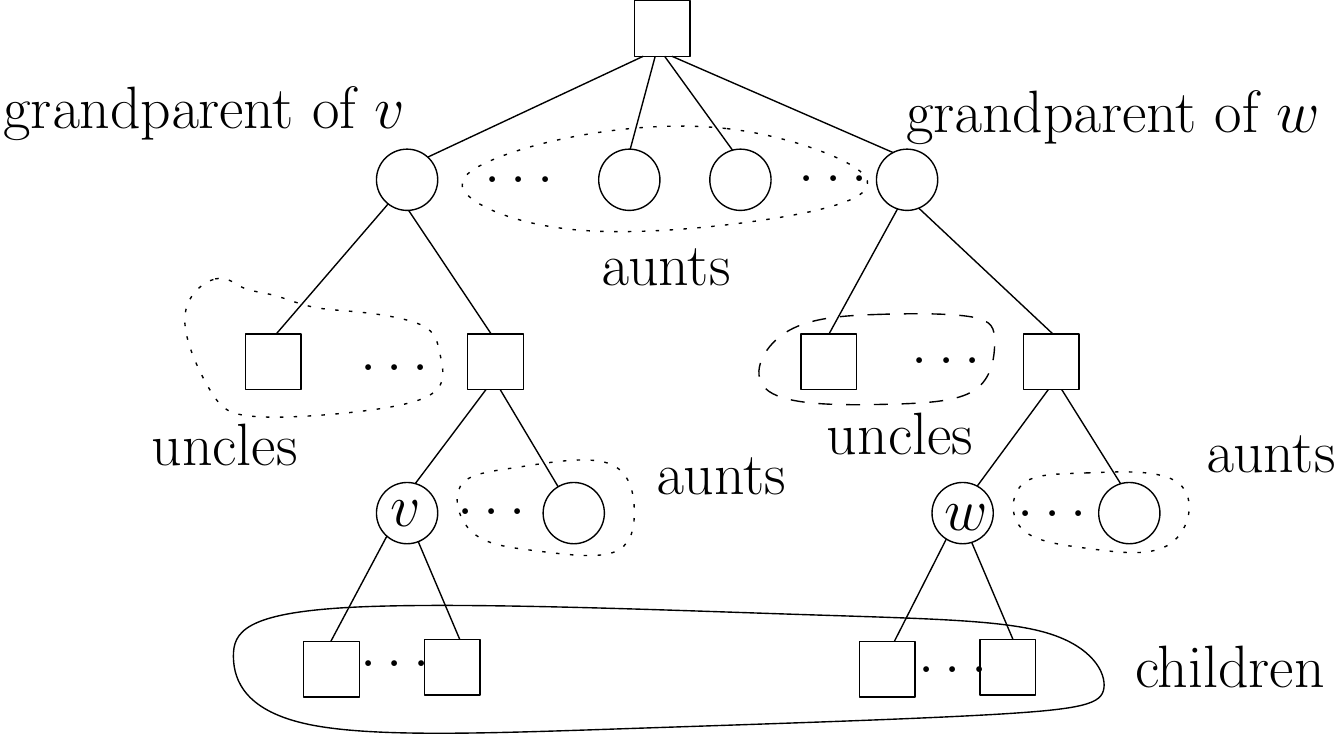}
\end{center}
Squares are the clause nodes and 
circles are the variable nodes. 
Note that we assume that
$v$ and $w$ are both on level 3, and their lowest common ancestor is the root. In the other cases, the picture and the subsequent calculation will be slightly different. 
To ease notation, we adopt the 
notation $\cut_p(u) := \cut_p(T_u)$,
where $T_u$ is the subtree of $T'$
rooted at $u$ (note that $T'$ and $T$
have the same node set, only some 
labels differ).
In the case depicted in the figure, 
$\tau$ is pivotal if and only if
\begin{enumerate}
    \item $\cut_p(u)$ happens 
    for all aunts and uncles $u$;
    \item $\cut_p(u)$ does not happen
    for all children $u$ of $v$;
    neither for all children $u$ of $w$.
    \item $\pi(\textnormal{grandparent of $v$}), \pi(\textnormal{grandparent of $w$}) \geq p$.
\end{enumerate}
Furthermore, note that
$\Pr[\cut_p(u)]$ equals $Q(p)$
if $u$ is an uncle and $R(p)$
if $u$ is an aunt. Therefore,
\begin{align*}
\Pr[\cut_p(T)] - \Pr[\cut_p(T')]
& \geq (p - p^2) \cdot 
    \Pr[\tau \textnormal{ is pivotal}] 
    = \\
    & (p - p^2)
    Q(p)^{\textnormal{uncles}}
    \cdot 
    R(p)^{\textnormal{aunts}}
    \cdot 
    \left(1 - Q(p)^{d-1} \right)^2 (1-p)^2 \\
    &
    =: \delta(p) \ .
\end{align*}
It is clear that $\delta(p) > 0$
for $0< p < 1 - 1/N$ and $\delta(p) = 0$
for $p \geq 1 - 1/N$.
Recalling the definition of 
$S_{d,k} = \E[\log(J_1 + \cdots + J_{d-1} + 1)]$ 
comparing it to $\E[\log_2(A_x)] = \E[\log_2(A_{x,1} + \dots + A_{x,d-1} + 1 )]$, we can couple the ensembles
$\mathbf{A} := (A_{x,c})_{c=1}^{d-1}$ and
$\mathbf{J} := (J_{c})_{c=1}^{d-1}$ such that $\mathbf{A} \leq \mathbf{J}$ except
with probability $\error(d,k,p,h)$, 
and $A_{x,c} = 0, J_c = 1$, conditioned
on $\pi(x) = p$, happens
with probability at least $\delta(p) - \error(d,k,p,h)$.
In fact, let us ignore the term 
$\error(d,k,h)$ for now and simply 
assume that $\mathbf{A} \leq \mathbf{J}$ 
(more rigorously, we would have to replace
every $T^h_{x,c}$ by the appropriate infinite
version; we decide to simply ignore
$\error(d,k,h)$ in the following, lest we 
overload the reader with our notation).
Set $\Delta := J - A_x$, and observe
that $\Delta \geq 0$ and 
$\Pr[\Delta \geq 1 \ | \ \pi(x)=p] \geq \delta(p)$. 
\begin{align*}
    \E[\log_2(J)]- \E [ \log_2 (A_x) ] & =
    -\E\left[ \log_2 \left(
    \frac{J-\Delta}{J} \right) \right]  \\
    & = 
    -\E\left[ \log_2 \left(1 - 
    \frac{\Delta}{J} \right) \right]\\ 
    & \geq 
    -\E\left[ \log_2 \left(1 - 
    \frac{\Delta}{d} \right) \right]\\ 
    & \geq 
   \frac{\log_2(e)}{d} \E\left[ 
    \Delta  \right]\\ 
    & \geq
    \frac{ \log_2(e)}{d}
    \int_0^1 \delta(p) \, dp =:
    \epsilon_{\rm privileged} \ .
\end{align*}
This is some positive number,
and it depends only on $d$ and $k$.
\end{proof}

\section{Local reasoning for ImpatientPPSZ}
\begin{lemma}[Lemma \ref{lemma-impatient-cut}, restated]
    \label{lemma-impatient-cut-appendix}
    Suppose $x\in \vbl(F)$ is non-priviledged.
        Let $p = \pi(x)$.
        If $\impcut_{p}(T^h_{x,c})$ happens then
        $A^{\imp}_{x,c} = 0$.
\end{lemma}
\begin{proof}
    We will prove the contrapositive: assume that 
    $A_{x,c}^{\imp} = 1$ and show that 
    $\impcut_p(T^h_{x,c})$ does not happen. 
    Let $F(T^h_{x,c})$ denote the set of clause labels
    appearing in $T^h_{x,c}$. 
    Since $A_{x,c}^{\imp} =1$ by assumption,
    the formula $F^{[V^\imp_x \mapsto d]}$ does not $D$-imply
    $(x \ne c)$. In particular, $|F(T^h_{x,c})| \leq D$ and 
    therefore $F(T^h_{x,c})^{[V^{\imp} \mapsto d]}$ does not 
    imply $(x \ne c)$. This means that there is an assignment
    $\gamma$ that (1) satisfies $F(T^h_{x,c})$, 
    (2) $\gamma(x) = c$, (3) $\gamma(y) = d$ for all
    $y \in V^{\imp}_x$.\\
    
    As a first step, we will show that
    $\cut_p(T^h_{x,c})$ does not happen. For this,
    we will construct a sequence of clause nodes $u_0, u_1, \dots$,
    with
    $u_0$ being the root andn $u_{i+1}$ being a grandchild 
    of $u_i$, keeping the following invariant:
    
    \begin{quotation}
        \textbf{Invariant.} For every clause node $u$ in the 
        sequence, $\beta_u(y) \ne d \Rightarrow \gamma(y) = \beta_u(y)$.
    \end{quotation}
    
    Note that the invariant is satisfied for the root:
    $x$ is the only variable with $\beta_{\root}(x) \ne d$,
    and $\gamma(x) = c = \beta_{\root}(x)$.
    To find $u_{i+1}$ from $u_i$, let $C_i$ be the clause
    label of $u_i$, and write $C_i$ as 
    \begin{align*}
        C_i = (y_1 \ne c_1 \vee \dots \vee y_l \ne c_l
        \vee z_{l+1} \ne d \vee \dots \vee z_{k-1} \ne d) \ ,
    \end{align*}
    where $c_1, \dots, c_l \ne d$. By construction,
    $\beta_{u_i}$ violates $C_i$, and therefore
    $\beta_{u_i}(y_j) = c_j$ for $1 \leq j \leq l$;
    by the invariant, $\gamma(y_j) = c_j$, too.
    But $\gamma$ satisfies $C_i$ (it satisfies
    every clause label in $T^h_{x,c}$), and therefore
    $\gamma(z_j) = c \ne d$ for some $l+1 \leq j \leq k-1$.
    In particular, $u_{i}$ has children.
    Let $v$ be the child of $u_i$ with variable
    label $z_j$. If $v$ is a leaf (a safe leaf), terminate
    the process and call the path from root to $v$
    the {\em witness path}.
    Otherwise, and let $u_{i+1}$ be the child of 
    $v$ with $EC(v, u_{i+1}) = c$. Note that $u_{i+1}$
    satisfies the invariant.\\
    
    Since $T^h_{x,c}$ is finite, this process terminates 
    with a witness path. Note that $\gamma(y) \ne d$
    for all variable labels $y$ appearing on that path. 
    In particular,
    this means that $y \not \in V^{\imp}_x$,
    thus $y \not \in V_x$, thus $\pi(y) \geq \pi(x)$.
    In other words, $\cut_p(T^h_{x,c})$ does not happen.\\
    
    Without loss of generality, let $v_1$ be the 
    level-1-node on the witness path, and $T_1$ be the tree
    rooted at $v_1$, and $y_1 := \varlabel(v_1)$. Observe that
    $\cut_p(T_1)$ does not happen. We will 
    now show that $\localimpcut_p(v_1)$ does not happen, either.
    Assume, for the sake of contradiction, that 
    $\localimpcut_p(v_1)$ happens. Does
    it happen because of Point~\ref{impcut-y-regular} in 
    the definition? Certainly not: $\gamma(y_1) \ne d$
    since $v_1$ is on the witness path, and thus
    $\pi(y_1) \geq p$. So it happens because of 
    Point~\ref{impcut-y-early}, and 
    $\eligible_{y_1} = 1$; without loss of generality,
    this means that $\localimpcut_p(v_1)$ happens
    for the first $d-2$ children $w_1,\dots,w_{d-2}$ 
    of $v_1$; let $C_1,\dots, C_{d-2}$ be the respective
    clause labels. All those $C_i$ are critical clauses
    ($x$ is non-priviledged, remember), and have $k-1$ 
    children each. So  $\localimpcut_p$ happens for 
    the first $(k-1)(d-2)$ of the $(k-1)(d-1)$
    grandchildren of $v_1$. In other words,
    all their variable labels $z$ have $\pi(z) < p$
    and thus $z \in V_x$. Under the assignment $[V_x \mapsto d]$,
    each of $C_i$ reduces to a unit clause;
    this unit clause is still violated by $\beta_{w_i}$
    and is therefore    either $(y_1 \ne i)$ 
    or $(x \ne c)$. If it was $(x \ne c)$ then
    $F(T^h_{x,c})^{[ V_x \mapsto d]}$ would imply $(x \ne c)$
    and therefore $A_{x,c} = A^{\imp}_{x,c}=0$, 
    contradicting our assumption. So it is $(y_1 \ne i)$.
    In other words, $F(T^h_{x,c})^{[V_x \mapsto d]}$ contains
    the unit clauses $(y_1 \ne 1), \dots, (y_1 \ne d-2)$;
    thus, when $x$ is being processed by ImpatientPPSZ,
    the set of plausible values for $y$ has been reduced
    to at most two values: $d-1$ and $d$; since 
    $\eligible_{y_1} = 1$, the algorithm will assign
    $y_1$ a value in Line~\ref{impAssigned}, and 
    $y_1 \in V^{\imp}_x$. This is again a contradiction:
    $\gamma(y_1) \ne d$ since $v_1$ is on the witness path;
    $\gamma(y_1) = d$ since $y_1 \in V^{\imp}_x$. This 
    concludes the proof.
\end{proof}
\section{ImpCut probability}
Suppose $x\in\vbl(F)$ is non-priviledged and $T^h_{x,c}$ is a critical clause tree for $x$ and $c\in[d]$.
\begin{lemma}[Lemma \ref{lemma-cut-prob-impatient}, restated]
\label{lemma-cut-prob-impatient-appendix}
  If $p < \theta$ then $  \Pr[\impcut_p(T^h_{x,c}) \ | \ \pi(x) = p]$ is at least
  \begin{align*}
      \left( p + c (\theta -p) \abamo( p^{k-1}, d-1)
       + (1 - p - c(\theta - p)) Q(p)^{d-1}
      \right)^{k-1} - \error(d,k,h) \ . 
  \end{align*}
  If $p \geq \theta$ then it is at least $Q(p)-\error(d,k,h)$.
\end{lemma}
\begin{proof}
If $p \geq \theta$ then this is obvious since 
already $\cut_p(T^h_{x,c})$ has probability at least 
$Q(p) - \error(d,k,h)$, by Proposition~\ref{lemma-cut-prob-Q}.
Thus we assume $p < \theta$. The root of $T^h_{x,c}$ has $k-1$ children $v_1,\dots,v_{k-1}$,
    whose respective variable labels are 
    $y_1,\dots,y_{k-1}$. Let $T_i$ denote
    the subtree of $T^h_{x,c}$ rooted at $v_i$.
\begin{align*}
  \Pr\left[\impcut_p(T^h_{x,c})\right] & =
  \Pr\left[\bigwedge_{i=1}^{k-1}
        \left(
         \cut_p(T_i)
         \vee
         \localimpcut_p(v_i)
         \right) 
         \right] \\  
         & \geq
    \prod_{i=1}^{k-1}
    \left(
         \Pr[\cut_p(T_i)
         \vee
         \localimpcut_p(v_i)]
         \right)
         \tag{FKG inequality} \ .
\end{align*}
We can apply the FKG inequality because 
each event $\cut_p(T_i)
         \vee
         \localimpcut_p(v_i)$
is a monotone increasing Boolean function
in the variables $[\pi(z) < p]$ and $\eligible_{y_i}$.
It remains to show that, for each $1 \leq i \leq k-1$, 
the event $
    \cut_p(T_i)
         \vee
         \localimpcut_p(v_i)
$
happens with probability at least     
\begin{align}
\label{formula-impcut-y}
         p + c (\theta -p) \abamo( p^{k-1}, d-1)
       + (1 - p - c(\theta - p)) Q(p)^{d-1} - \error(d,k,h)
\end{align}
For this, let us abbreviate $T := T_i$, $v := v_i$ its root,
and $y := \varlabel(v) = y_i$; also, we define
the events $A := \localimpcut_p(v)$
and $B := \cut_p(T)$. 
We distinguish three cases: 

\begin{enumerate}
    \item[(i)] if (1) $\pi(y) < p$ then 
the desired event $A \vee B$ happens;
    \item[(ii)] if $\pi(y) \geq p$ and $\eligible_y=1$ (which
    implies $\pi(y) < \theta$) then
    we ignore $B$ and focus on $A$;
    \item[(iii)] if $\pi(y) \geq p$ and $\eligible_y=0$,
    then $A$ does not happen, so focus on $B$.
\end{enumerate}

Formally,
\begin{align*}
    \Pr[A \vee B] & \geq \Pr[(i)] 
    + \Pr[(ii)]
    \cdot \Pr[A \ | \ (ii)]
    + 
    \Pr[(iii)] \cdot 
    \Pr[B \ | \ (iii)]
\end{align*}
Next, let us look at each case.
\begin{enumerate}
    \item $\Pr[(i)] = p$; this explains the first
    term in (\ref{formula-impcut-y}).
    \item $\Pr[(ii)] = c (\theta - p)$. Furthermore, if
     if (ii) happens, then $A$ happens
if and only if for at least $d-2$ of the children $w_1,\dots,w_{d-1}$,
the event $A_j := \localimpcut_p(w_j)$ happens.
Each $A_j$ happens with probability $\rho := p^{k-1}$; they are 
independent since all $(d-1)(k-1)$ grandchildren of $v$
have distinct labels. Therefore,
\begin{align*}
    \Pr[A \ | \ (ii)]  & = 
    \Pr[A_1 \wedge \dots \wedge A_{d-1}] + 
    \sum_{j^* = 1}^{d-1}
 \Pr[\neg A_{j^*} \wedge \bigwedge_{j \ne j^*} A_j] \\
    & = 
    \rho^{d-1} + (d-1) (1-\rho) \rho^{d-2}
    = \abamo(p^{k-1}, d-1) \ .
 \end{align*}
     This explains the second term in (\ref{formula-impcut-y}).
     \item $\Pr[(iii)] = 1 - p - c(\theta  -p)$.
     If (iii) happens, then $B$ happens if and only
     if $\cut_p(T')$ happens for each of the $d-1$ 
     subtrees of $T$. By Proposition~\ref{lemma-cut-prob-Q},
     this happens with probability
     $\left( Q(p) - \error(d,k,h) \right)^{d-1}$.
     This explains the third and fourth term
     in (\ref{formula-impcut-y}).
\end{enumerate}
This concludes the proof.
\end{proof}

\section{Bounding losses and gains. Proofs of 
Propositions~\ref{proposition-loss} and~\ref{proposition-gain}}
\label{section-appendix-loss-gain}

First, we need some good-enough estimates 
for our probabilities $R(p)$, $Q(p)$, and $W(p)$.
Note that $R(p)$ and $Q(p)$ are the roots of certain polynomials,
and we do not have an explicit formula for them.
The bounds in Proposition~\ref{proposition-RWQ-bound-appendix} are somewhat crude but sufficient for our purposes.

\begin{proposition}
\label{proposition-RWQ-bound-appendix}
    $R(p) \leq p + 4\, p^L$;
    $Q(p) \leq \left(p + 4\, p^L \right)^{k-1}$;
    and
    $W(p) \leq p + O(\theta p^{(d-2)(k-1)})$.
    The hidden constant in the $O$ depends
    on $d$ and $k$ only.
\end{proposition}

\begin{proof}
One checks that $R(p)$ is convex on the 
interval $[0, 1 - 1/L]$. To see this, note that
for $p \leq 1 - 1/L$,
   $R(p)$ is the unique solution in $[0,1]$
   of the equation
   \begin{align*}
       R = p + (1-p) R^{L} \ ,
   \end{align*}
   by Proposition~\ref{ExtinProbPPSZ}.
   We can solve explicitly for $p$
   and check that $p(R)$ is concave, 
   by elementary calculus.
   Since $R$ is convex, $R(0) = 0$,
   and $R(1 - 1/L) = 1$, the graph of $R(p)$
   is below the line from $(0,0)$
   to $(1 - 1/L,1)$, and therefore
   $R(p) \leq \frac{L}{L-1} p$.
   This is 
   not enough yet, but 
   applying the equation of $R$ 
   to this estimate gives 
   \begin{align*}
       R = p + (1-p) R^{L} 
       \leq p + (1-p) \left( \frac{L}{L-1}\, p\right)^{L}
       \leq p + 4\, p^L \ .
   \end{align*}
   The upper bound for $Q$ follows directly
   from $Q(p) = R(p)^{k-1}$. It remains 
   to prove the upper bound on $W(p)$:
   \begin{align*}
       W(p) & = p + c(\theta -p) \abamo(p^{k-1}, d-1)
       +  (1 - p - c(\theta - p)) Q(p)^{d-1} \\
       & \leq p + \theta \abamo(p^{k-1}, d-1)
       + Q(p)^{d-1} \\
       & =
       p +  \theta p^{(k-1)(d-1)} + \theta (d-1) (1-p^{k-1})p^{(k-1)(d-2)} + Q(p)^{d-1} \\
       & \leq
       p +(d-1) \theta p^{(d-2)(k-1)} 
       + R^L \\
       & \leq p + (d-1) \theta p^{(d-2)(k-1)}
       + (p + 4\, p^L )^L \\
       & \leq p + O\left( \theta p^{(d-2)(k-1)} \right)\  .
   \end{align*}
\end{proof}

\begin{proposition}[Proposition~\ref{proposition-loss}, restated]
\label{proposition-loss-appendix}
   $\Pr[ J^{\imp} = 1 \wedge \eligible_x]  \leq 
   \frac{c}{L+1} \theta^{L+1}
   + O\left( \theta^{L+2} \right)$.
\end{proposition}
\begin{proof}
   Recall that if $\pi(x) < \theta$ then 
   $\eligible_x$ is $1$ with probability $c$ and $0$
   with probability $1-c$. If $\pi(x) \geq \theta$
then    $\eligible_x = 0$.   
Also, $J^{\imp} = 1$ if and only if 
$J^{\imp}_{1} = \dots = J^{\imp}_{d-1} = 0$.
   Therefore,
   \begin{align*}
       \Pr[ J^{\imp} = 1 \wedge \eligible_x] & = 
       c \cdot \int_0^{\theta} \Pr[J^{\imp} = 1 \ | \ \pi(x) = p]
       \, dp 
        =
       c \cdot \int_0^{\theta} W^{(d-1)(k-1)} \, dp \\
       & = 
       c \cdot \int_0^{\theta} (p + O(\theta p^{(d-2)(k-1)}) )^{L} \, dp
       \leq 
       c \cdot \int_0^{\theta}
        p^L (1  +O(\theta) ) \, dp \tag{since $(d-2)(k-1) \geq 1$} \\
        & = \frac{c}{L+1} \theta^{L+1} + O\left( \theta^{L+2} \right)
   \end{align*}
   This proves the proposition.
\end{proof}

\begin{proposition}[Proposition~\ref{proposition-gain}, restated]
\label{proposition-gain-appendix}
    $\E\left[\log_2  (J^{\imp})\right] - S_{d,k} \leq
    (d-1) \log_2 (1 - 1/d) \cdot 
    \left(
    \frac{c}{L+1} \theta^{L+1} + O\left( \theta^{L+2} \right)
    \right)
    $.
\end{proposition}

\begin{proof}
Recall the definition of $S_{d,k}$:
sample random variables $J_1, \dots, J_{d-1}$ 
by setting $p := \pi(x)$ and setting each
$J_c$ to $0$ with probability $Q(p)$
and to $1$ with probability $1 - Q(p)$, and $J = J_1 + \cdots + J_{d-1} + 1$. So the $J_c$ are independent conditioned
on $\pi(x) = p$. Then
$S_{d,k} = \E[ \log_2(J)]$. 
Set $\Delta_c := J_c - J^{\imp}_c$ and $\Delta = 
\sum_c \Delta_c$. Note that all $\Delta_c$ have the same distribution.

\begin{proposition}
\label{prop-Delta-1}
$    \E[\Delta_1 \ | \ \pi(x) = p] \geq
    c (\theta - p) L \left( p^{L-1} - O(p^L) \right)$ for all $1 \leq c \leq d-1$. 
\end{proposition}

In particular, if $p < \theta$ and $\theta$ is sufficiently small then
$\E[\Delta_1] \geq 0$. Therefore, $\E[J_c] \leq \E[J^{\imp}_c]$ and we can couple 
the ensemble $(J_1,\dots,J_{d-1})$ and $(J^{\imp}_1,\dots,J^{\imp}_{d-1})$
on a common probability space on which $J_c \leq J^{\imp}_c$, always, and thus $\Delta \geq 0$.
We therefore see that $\E\left[ \log_2( J^{\imp})\right] - S_{d,k}$ is 
\begin{align*}
&
    \E[\left[ \log_2(J^{\imp}) - \log_2(J)\right] 
 = 
    \E\left[ \log_2 \left(1 - \frac{\Delta}{J}\right) \right] \\
     & \leq 
    \E\left[ \log_2 \left(1 - \frac{\Delta}{d}\right) \right] 
    \leq 
    \E\left[ \log_2 \left( \left(1 - \frac{1}{d} \right)^{\Delta} \right) \right] \\
    & =
    \E[\Delta] \log_2 \left(1 - \frac{1}{d} \right) \ . 
\end{align*}
Conditioned on $\pi(x) = p$ and using Proposition~\ref{prop-Delta-1}, this is at most
\begin{align*}
    c (\theta - p) L \left( p^{L-1} - O(p^L) \right)  (d-1)  \log_2 \left(1 - \frac{1}{d} \right) \ .
\end{align*}
We integrate this over $p$ to get rid of the condition $\pi(x) = p$ and see that 
\begin{align*}
    \E\left[\log_2(J^{\imp})\right] - S_{d,k}    
    &  \leq (d-1)\log_2\left(1 - \frac{1}{d} \right)
     \cdot \left( \frac{c}{L+1} \theta^{L+1} + O\left( \theta^{L+2}\right) 
     \right) \ . 
\end{align*}
This concludes the proof of Proposition~\ref{proposition-gain-appendix}.
\end{proof}

It remains to prove Proposition~\ref{prop-Delta-1}.

\begin{proof}[Proof of Proposition~\ref{prop-Delta-1}].
\begin{align*}
    \E[\Delta_1 \ | \ \pi(x) = p] & = \E[J_c - J_c^{\imp} \ | \ \pi(x) = p]
    &= 
    (1 - Q) - (1 - W^{k-1}) = W^{k-1} - R^{k-1} \\
    & \geq(k-1) (W-R) R^{k-2} \ ,
\end{align*}
where the last inequality follows because
$W^{k-1} = (R + W - R)^{k-1} = R^{k-1} \left(1 + \frac{W-R}{R} \right)^{k-1} \geq R^{k-1} \left(1 + \frac{(k-1)(W-R)}{R} \right) = R^{k-1} + (k-1)(W-R)R^{k-2}$.
Now let us bound $W-R$ from below. If $p \geq \theta$
then $W(p) = R(p)$ and $W-R = 0$. If $p < \theta$, we expand
$R(p)$ as follows:
\begin{align*}
    R = p  + (1-p) Q^{(d-1)}
    = p + c(\theta - p) Q^{d-1} + (1 - p - c(\theta-p)) Q^{d-1}
\end{align*}
and therefore
\begin{align*}
    W - R & = c (\theta - p) \left( \abamo (p^{k-1}, d-1) - Q^{d-1} \right)  \\
    & = c(\theta - p) \left(
    p^{(k-1)(d-1)} + (d-1) \left(1 - p^{k-1} \right)
    p^{(k-1)(d-2)} - Q^{d-1} \right) \\
    & \geq
    c (\theta - p) \left( p^L + (d-1) p^{(k-1)(d-2)}
    - (d-1) p^L - ( p + O(p^2))^{L} \right) \\
    & \geq
    c (\theta - p) (d-1) \left( p^{(k-1)(d-2)} - O(p^L) \right)
     \ .
\end{align*}
Next, combining the previous two calculations, we see that
\begin{align*}
    \E[\Delta_1 \ | \ \pi(x) = p] & \geq(k-1)(W-R) R^{k-2}
 \geq (k-1)(W-R) p^{k-2}\\
    & \geq (k-1) c (\theta - p) (d-1) \left(
    p^{(k-1)(d-2)} - O(p^L)
    \right) p^{k-2}\\
    & \geq
    c (\theta - p) L \left( p^{L-1} - O(p^L) \right) \ . 
\end{align*}

\end{proof}

\end{document}